\documentstyle[aps,preprint,epsf]{revtex} 
\newcommand{\inst}[1]{$^{#1}$} 
\newcommand{\stars}{\begin{center}***\end{center}} 
\begin{document} 
%*******************************************
\draft
%*******************************************
\title{Beyond the Fokker-Planck equation: Stochastic Simulation 
of Complete Wigner representation for 
the Optical Parametric Oscillator} 
%*******************************************
\author{L.\ I.\ Plimak\inst{1,2,3}, M.\ K.\ Olsen\inst{1,4}, 
M.\ Fleischhauer\inst{3} 
and M.\ J.\ Collett\inst{1}} 
%*******************************************
\address{  
\inst{1}Department of Physics, University of Auckland, Private Bag 
92019, Auckland, New Zealand\\  
\inst{2}Department of Chemical Physics, The Weizmann 
Institute of Science, 76100 Rechovot, Israel\\  
\inst{3}Fachbereich Physik, Universit\"at Kaiserslautern, 
D-67663 Kaiserslautern, Germany\\  
\inst{4}Instituto de F\'{\i}sica, Universidade Federal Fluminense, 
Av. Litor\^anea s/n, Boa Viagem CEP 24210-340, Niter\'oi - RJ, Brazil 
}
%*******************************************
%\date{\today}
%*******************************************
\maketitle
%******************************************* 
\begin{abstract} 
%******************************************* 
We demonstrate a method which allows the stochastic 
modelling of quantum systems for which the 
generalised Fokker-Planck equation in the phase space contains 
derivatives of higher than second order. 
This generalises quantum stochastics far beyond the 
quantum-optical paradigm of 
three and four-wave mixing problems to which these techniques 
have so far only been applicable. 
To verify our method, 
we model a full Wigner representation for the optical 
parametric oscillator, 
a system where the correct results are well known and can 
be obtained by other methods. 
%******************************************* 
\end{abstract} 
\newpage
%******************************************* 
\section{Introduction} 
%******************************************* 
Coherent-state, or phase-space, path integrals 
\cite{Klauder} 
are so far the only variety of path integrals 
that can be evaluated numerically 
for real time problems. 
For interactions 
which are no more than quadratic in creation or 
annihilation operators, the measure over the paths 
can be described in terms of 
stochastic differential equations (SDEs, Langevin equations) for 
the paths \cite{Crispin,stocmeth}. 
Unlike, e.g., the Feynman path integral, 
such real-time phase-space path integrals 
are characterised by a constructively 
defined positive measure, and can be 
calculated by 
simulating the corresponding Langevin equations. 
These techniques 
have been most successfully used in quantum optics 
\cite{Crispin}, 
but have also been applied to the quantum dynamics 
of condensed bosonic atoms \cite{Steel}. 
Mathematically, however, the 
existence of an SDE for the paths 
is restricted to problems were the equation for 
the corresponding pseudo-probability distribution 
is a genuine Fokker-Planck equation 
(this is the content of Pawula's theorem~\cite{Pawula}). 
This imposes the above restriction on the 
interaction Hamiltonian and thus confines 
the class of problems for which the 
measure of the path integral may 
be characterised in terms of an SDE 
to three and four-wave mixing problems. 
(The two-body collisional interaction of bosons 
has the formal structure of four wave mixing.) 

There exist, however, a wide variety of problems where the 
generalised Fokker-Planck equation is of third or 
higher order and hence 
Langevin equations may not be derived. 
As an example, a description of nonlinear quantum optical 
processes in terms of the Wigner 
distribution results in generalised Fokker-Planck 
equations with third-order derivatives~\cite{stocmeth}. 
A common procedure is to truncate the Wigner 
equation at second order, which is equivalent to using the 
semiclassical theory of stochastic electrodynamics~\cite{trevor}. 
This procedure necessarily discards 
the deeper quantum aspects of the 
problem and gives answers at odds 
with quantum mechanics for several 
systems~\cite{BadTrWig}. 
There are also situations 
where even a P-representation Fokker-Planck 
equation must also be written in generalised form~\cite{cavBEC}, and more are likely to be 
investigated in the future. 
It is therefore of considerable interest to generalise methods 
allowing for a constructive characterisation of the measure of the 
phase-space path integral beyond the very 
restrictive quantum-optical paradigm. 

Pawula's theorem would at first 
sight seem to forbid this generalisation. 
However, on closer 
inspection we see that the restrictions 
imposed can be evaded by discretising time. 
Full technical details are 
presented elsewhere \cite{BWO}; 
the aim of this letter is to demonstrate the feasibility of 
this generalisation, using as a demonstrative example 
the Wigner representation for the 
optical parametric oscillator (OPO). 
We chose this system because it is well known and the 
results can be obtained by other methods. 
We derive a system of stochastic difference equations 
in a doubled phase-space, 
which is related to the Wigner representation 
in the same way as the well-known positive-P 
equations~\cite{JPAP+} are related to the P-representation. 
By analogy, we shall call this representation 
the positive-W representation. 
We show that this representation 
gives the correct results for quadrature relaxation, 
whereas the truncated Wigner representation makes noticeably 
different predictions. 

It should be stressed that 
no continuous time limit exists for the positive-W 
equations. 
Unlike the Wiener process, where sampling noise is independent 
of the time step for a given sample size, 
here sampling noise diverges in the continuous time 
limit. 
This is how Pawula's theorem is enforced. 
In practice, however, one is interested in the sampling noise 
{\em vs\/} time step not for a given sample size, but for a given 
computational time. 
From this perspective, the difference we find is much less 
dramatic. 
Although, as the time step 
is decreased, the computational 
time grows at a faster rate than 
for conventional stochastic integration, 
this dependence remains polynomial so that 
the problem stays in the same class of computational 
complexity.  

%******************************************* 
\section{Positive-W representation} 
%******************************************* 
Degenerate optical parametric oscillation is an optical process 
in which 
a nonlinear medium inside an optical cavity is pumped with light at 
one frequency and emits light at half that frequency. 
The free, damping and pumping Hamiltonians for this system may be written 
in their usual 
forms~\cite{DFW} while we 
choose the nonlinear coupling constant between the light modes, $\kappa$, 
to be real so that 
the interaction Hamiltonian is 
\begin{equation} 
H_{%
{\rm int}%
} = \frac{i\hbar\kappa }{2}\left[ \hat a^{\dag\;2}\hat b 
- \hat a^{2}\hat b^{\dag}\right]. 
\label{eq:ham} 
\end{equation}%
The annihilation (creation) operators $\hat{a}\;(\hat{a}^{\dag})$ and 
$\hat{b}\;(\hat{b}^{\dag})$ annihilate (create) photons at the lower 
and higher frequencies respectively. 
Proceeding via the usual methods, we can map the Hamiltonian onto 
differential equations for the Wigner and positive-P 
distributions~\cite{Crispin}. The positive-P representation gives 
a Fokker-Planck equation and can thus be mapped straightforwardly 
onto a set of coupled SDEs using It\^o 
rules. The equation of motion for the Wigner distribution, however, 
has third-order derivatives 
and thus has no mapping onto stochastic differential 
equations. 

The positive-P representation can alternatively be 
derived~\cite{BWO} 
by postulating a mapping of time-normal ($T_N$) 
averages~\cite{Roy} of 
the Heisenberg quantum-field operators, 
$\hat{\sl a}%
^{\dag}%
(t), 
\hat{\sl b}%
^{\dag}%
(t), 
\hat{\sl a}(t)$, 
$\hat{\sl b}(t) 
$, onto classical averages. 
For example, 
\begin{eqnarray} % \nonumber 
\label{eq:Four} 
\left\langle 
T_N\left\{\hat{\sl a}%
^{\dag}%
(t) 
\hat{\sl b}%
^{\dag}%
(t')\hat{\sl a}(t'') 
\hat{\sl b}(t''')\right\}%
\right\rangle 
&\equiv& 
\left\langle  
T_- 
\left \{ 
\hat{\sl a}%
^{\dag}%
(t) 
\hat{\sl b}%
^{\dag}%
(t') 
\right \} 
T_+ 
\left \{ 
\hat{\sl a}(t'') 
\hat{\sl b}(t''') 
\right \} 
\right\rangle%
\nonumber\\ 
&=& 
\overline{\hspace{0.1ex} 
{a}%
^{\dag}%
(t) 
{b}%
^{\dag}%
(t') 
{a}(t'') 
{b}(t''') 
\hspace{0.1ex}} 
\ 
. 
\end{eqnarray}%
The upper bar on the RHS of this relation denotes averaging 
over the statistics of the four stochastic c-number fields, 
${a}%
^{\dag}%
(t), 
{b}%
^{\dag}%
(t), 
{a}(t), 
{b}(t) 
$ (i.e. phase-space path integration). 
The essence of the positive-P representation is in defining 
these statistics constructively. 
The Heisenberg equations of motion are mapped onto 
the SDE's for the fields, 
while the averaging over the initial state of the 
quantum field (denoted as $%
\left\langle 
\cdots%
\right\rangle%
$) 
is mapped onto the distribution over the initial conditions. 

The $T_N$ ordering emerges when we generalise the normal ordering 
of free-field operators to the Heisenberg operators. 
Generalising the symmetric, or Wigner, 
ordering of free-field operators results in 
a {\em time-Wigner\/} ($T_W$) ordering of the Heisenberg 
field operators~\cite{BWO}. 
Under $T_N$, the ``most recent'' creation (annihilation) operator 
becomes the leftmost (rightmost) in the product. 
The $T_W$-ordering acts in a similar way except 
that it is 
symmetrised with respect to the creation 
and annihilation 
operators: 
\begin{eqnarray} 
T_W \left\{ 
\hat {\sl x}(t)\hat {\sl y}(t')\cdots\hat {\sl z}(t'') 
\right\} &=& 
\frac{1}{2}\left[ 
T_W \left\{ 
\hat {\sl y}(t')\cdots\hat {\sl z}(t'') 
\right\}\hat {\sl x}(t) 
\right. \nonumber \\ 
&& + \, \left. 
\hat {\sl x}(t) T_W \left\{ 
\hat {\sl y}(t')\cdots\hat {\sl z}(t'') 
\right\} 
\right] 
. 
\label{eq:TWDef} 
\end{eqnarray}%
Here, $\hat {\sl x},\hat {\sl y},\cdots,\hat {\sl z}$ 
stand for field operators 
(i.e. $\hat {\sl a}$ or $\hat {\sl a}%
^{\dag}%
$), 
and $t$ should exceed 
all other time arguments in the product so that 
$t>t',\cdots,t''$. 
Equation (\ref{eq:TWDef}) is a recurrence relation which 
defines $T_W$ for the case of different time arguments. 
The full definition may be found in~\cite{BWO}. 

Following the example of the positive-P mapping 
(\ref{eq:Four}), 
we postulate a {\em positive-W} mapping relating 
$T_W$-ordered operator averages to classical averages 
of four stochastic 
c-number fields, 
$\alpha (t),\alpha 
^{\dag}%
(t),\beta (t),\beta 
^{\dag}%
(t)$, 
so that, e.g., 
\begin{eqnarray} 
\left\langle  
T_W 
\left\{ 
\hat{\sl a}%
^{\dag}%
(t) 
\hat{\sl b}%
^{\dag}%
(t') 
\hat{\sl a}(t'') 
\hat{\sl b}(t''') 
\right\} 
\right\rangle 
= 
\overline{\hspace{0.1ex} 
{\alpha }%
^{\dag}%
(t) 
{\beta}%
^{\dag}%
(t') 
{\alpha}(t'') 
{\beta}(t''') 
\hspace{0.1ex}} 
\label{eq:HFWM} 
. 
\end{eqnarray}%
We emphasise that this mapping is distinct from that of 
Eq.\ (\protect\ref{eq:Four}) 
by changing the notation of the 
c-number fields. 
In full detail, the quantum-field-theoretical (QFT) 
techniques which we used in order to characterise 
the mapping (\ref{eq:HFWM}) constructively are 
described elsewhere \cite{BWO,ClDiag}. 
For the purposes of this letter, we note that these are a 
straightforward adaptation of similar QFT techniques 
described in Refs.\ \cite{Imag,Florence}. 
In \cite{Imag}, Matsubara-style quantum dynamics were 
mapped onto an imaginary-time SDE; 
in \cite{Florence}, Feynman diagram techniques were 
mapped onto a real-time SDE. 
Here, we consider dynamics on the Schwinger-Keldysh 
C-contour~\cite{Keldysh}. 
To include the Wigner representation, one needs a generalisation of 
Wick's theorem to the case of symmetric ordering. 
This generalisation is quite straightforward and 
results in a Keldysh-style diagram series for 
the $T_W$-ordered averages. 
As in \cite{Imag,Florence}, 
propagators in this diagram series are then expressed 
by the {\it retarded} Green's 
function of the free Schr\"odinger equation, 
and the whole series is restructured so as to make 
this Green's function a propagator in a new series. 
This yields a Wyld-type series~\cite{Wyld}, 
also termed {\em causal series\/}~\cite{ClDiag}. 
By applying multiple Hubbard-Stratonovich 
transformations~\cite{HST}, 
we eventually arrive at a classical stochastic problem 
for which this series is a solution. 
This final step of the derivation is of independent 
interest and is discussed in more detail below. 

In the strict mathematical sense, 
this derivation fails: 
no continuous-time process exists satisfying 
Eq.\ (\protect\ref{eq:HFWM})%
. 
The way around this problem is to 
allow the mapping (\ref{eq:HFWM}) 
to hold only approximately, and consider stochastic 
processes in discretised time. 
We then find 
the following set of stochastic 
{\em difference\/} equations, 
(dropping the $t$-dependence for notational simplicity) 
\begin{eqnarray} 
\Delta\alpha&=&\left(-\gamma_{1}\alpha+ 
\kappa\alpha^{\dag}\beta\right)\Delta t+ 
\sqrt{\gamma_{1}}\eta_{1}\Delta t^{1/2} 
+\sigma_{1}\Delta t^{1/3}, 
\nonumber 
\\ 
\Delta\alpha^{\dag}&=&\left(-\gamma_{1}\alpha^{\dag}+\kappa\alpha 
\beta^{\dag}\right)\Delta t+ 
\sqrt{\gamma_{1}}\eta_{1}^{\ast}\Delta t^{1/2}+\sigma_{1}^{\dag}\Delta t^{1/3},\nonumber\\ 
\Delta\beta&=&\left(\epsilon-\gamma_{2}\beta- 
\frac{\kappa}{2}\alpha^{2}\right)\Delta t 
+\sqrt{\gamma_{2}}\eta_{2}\Delta t^{1/2}+\sigma_{2}\Delta t^{1/3} 
,\nonumber\\ 
\Delta\beta^{\dag}&=&\left(\epsilon^{\ast}-\gamma_{2}\beta^{\dag}- 
\frac{\kappa}{2}\alpha^{\dag\;2}\right)\Delta t 
+\sqrt{\gamma_{2}}\eta_{2}^{\ast}\Delta t^{1/2}+ 
\sigma_{2}^{\dag}\Delta t^{1/3}, 
\label{eq:FD} 
\end{eqnarray} 
where 
$\Delta t$ is the step of time discretisation, 
$ 
\Delta\alpha(t)=\alpha(t+\Delta t)-\alpha(t) 
$ 
(likewise for the other field variables) and 
$\eta _{1,2}(t)$ are independent complex standardised Gaussian noises 
such that 
\begin{eqnarray} % \nonumber % \eqlabel{} 
\overline{\hspace{0.1ex}%
\eta _1(t)%
\hspace{0.1ex}} 
= 
\overline{\hspace{0.1ex}%
\eta _2(t)%
\hspace{0.1ex}} 
= 
0 
, \ \ 
\overline{\hspace{0.1ex}%
\eta _1(t)\eta _1(t')%
\hspace{0.1ex}} 
= 
\overline{\hspace{0.1ex}%
\eta _2(t)\eta _2(t')%
\hspace{0.1ex}} 
= 
0 
, \ \ 
\overline{\hspace{0.1ex}%
\eta _1(t)\eta _1^*(t')%
\hspace{0.1ex}} 
= 
\overline{\hspace{0.1ex}%
\eta _2(t)\eta _2^*(t')%
\hspace{0.1ex}} 
= 
\delta _{tt'} 
. 
\end{eqnarray}%
It should be noted that the $\eta$'s are $\delta _{tt'}$ 
(Kronecker) correlated, not $\delta (t-t')$ 
(Dirac) correlated and that we have explicitly taken care 
of the proper powers of $\Delta t$. 
The $\gamma_{j},\ j=1,2$, are the cavity loss rates at each frequency and $\epsilon$ 
represents the classical pump. 
For the $\sigma $'s we have 
\begin{eqnarray} % \nonumber % \eqlabel{} 
\sigma_1 = q\xi _2 
+ s\,\xi _1^{\dag *}\sqrt{p^{\dag} \xi _2^{*}} 
\,,\ \ 
\sigma_1^{\dag} = q^{\dag}\xi _2^{\dag} 
+ s^{\dag}\,\xi _1^{*}\sqrt{p \xi _2^{\dag *}} 
\,,\ \ 
\sigma_2 = r \xi _1 \sqrt{p \xi _2^{\dag *}} 
\,,\ \ 
\sigma_2^{\dag} = 
r^{\dag} \xi^{\dag} _1 \sqrt{p^{\dag} \xi _2^{*}}\,, 
\end{eqnarray}%
where $\xi _1,\xi _1^{\dag},\xi _2,\xi _2^{\dag}$ are 
independent complex standardised Gaussian noises 
(with the same properties as $\eta _1,\eta _2$). 
The other parameters obey the relations 
\begin{eqnarray} % \nonumber 
\label{eq:Pars} 
pq^{\dag} = p^{\dag}q = -\frac{\kappa }{8}, \ \ 
rs^{\dag} = r^{\dag}s = 1. 
\end{eqnarray}%
Within these constraints they can be chosen at will and 
may in fact even be field and/or time-dependent. 
This freedom can be used to control sampling noise in simulations. 

Comparing equations (\ref{eq:FD}) to 
the partial differential equation for the $W$-function, 
\begin{eqnarray} 
\frac{\partial W(\alpha,\beta,t)}{\partial 
t}&=&\left[\frac{\partial}{\partial 
\alpha}\left(\gamma_{1}\alpha-\kappa\alpha^{\ast}\beta\right) 
+\frac{\partial}{\partial\alpha^{\ast}}\left(\gamma_{1} 
\alpha^{\ast}-\kappa\alpha\beta^{\ast}\right)\right.\nonumber\\ 
& &\left.+\frac{\partial}{\partial\beta}\left(\gamma_{2}\beta+ 
\frac{\kappa}{2}\alpha^{2}-\epsilon\right)+\frac{\partial} 
{\partial\beta^{\ast}}\left(\gamma_{2} 
\beta^{\ast}+ 
\frac{\kappa}{2}\alpha^{\ast\;2}-\epsilon^{\ast}\right)\right.\nonumber\\ 
& &\left.+\left(\gamma_{1}\frac{\partial^{2}} 
{\partial\alpha\partial\alpha^{\ast}}+\gamma_{2}\frac{\partial^{2}} 
{\partial\beta\partial\beta^{\ast}}\right)\right.\nonumber\\ 
& &\left.+\frac{\kappa}{8}\left(\frac{\partial^{3}} 
{\partial\alpha^{2}\partial\beta^{\ast}}+ 
\frac{\partial^{3}}{\partial\alpha^{\ast\;2} 
\partial\beta} \right)\right]W(\alpha,\beta,t), 
\label{eq:WigFP} 
\end{eqnarray} 
we see that there is an obvious one-to-one correspondence between 
$n$-th order derivatives in (\ref{eq:WigFP}) and 
terms $\propto\Delta t^{1/n}$ on the RHS's of 
Eqs.\ (\protect\ref{eq:FD})%
. 
More precisely stated, each term in (\ref{eq:WigFP}) 
corresponds to 
a particular {\em cumulant\/}~\cite{Risken} 
of increments in (\ref{eq:FD}). 
This means that, for example, $\frac{\partial}{\partial 
\alpha}\left(\gamma_{1}\alpha-\kappa\alpha^{\ast}\beta\right)W$ 
specifies $%
\overline{%
\overline{\hspace{0.1ex}%
\Delta \alpha 
\hspace{0.1ex}}%
} 
= \left(-\gamma_{1} 
\alpha+\kappa\alpha^{\ast}\beta\right)\Delta t$, 
resulting in the contribution 
$\left(-\gamma_{1}\alpha+\kappa\alpha^{\ast}\beta\right)\Delta t$ 
to $\Delta \alpha $. Note that the double upper bar is used as a notation 
to signify cumulants. 
We can now also see that $\gamma_{1}\frac{\partial^{2}} 
{\partial\alpha\partial\alpha^{\ast}}W$ specifies 
$%
\overline{%
\overline{\hspace{0.1ex}%
\Delta \alpha \Delta \alpha 
^{\dag}%
\hspace{0.1ex}}%
} 
= \gamma _1 \Delta t$ 
and thus 
yields the contributions $\eta _1 \sqrt{\gamma _1\Delta t}$ 
to $\Delta \alpha $ and 
$\eta _1^* \sqrt{\gamma _1\Delta t}$ to $\Delta \alpha 
^{\dag}%
$. 
Finally, the third order derivatives 
are represented by the $\sigma$'s, with 
the latter defined in such a way as to have only two 
nonzero third-order cumulants of the increments, 
\begin{eqnarray} % \nonumber 
\label{eq:Cu4} 
\overline{%
\overline{\hspace{0.1ex}%
\Delta \alpha ^2\Delta \beta 
^{\dag}%
\hspace{0.1ex}}%
} 
= 
\overline{%
\overline{\hspace{0.1ex}%
\Delta \alpha 
^{\dag 2}%
\Delta \beta 
\hspace{0.1ex}}%
} 
= 
-\frac{\kappa \Delta t}{4}. 
\end{eqnarray}%

The one-to-one correspondence between 
(\ref{eq:FD}) and (\ref{eq:WigFP}) suggests 
that rules may be devised for 
finding coefficients in the 
positive-W equations, starting from 
the equation for the W-function. 
(This would however not constitute their 
{\em derivation\/}: whereas 
the $W$-function applies only to same-time symmetrically 
ordered operator averages, 
(\ref{eq:HFWM}) and (\ref{eq:FD}) cover a much wider 
class of multi-time, time-Wigner ordered averages.) 
QFT methods may then be of much assistance when 
factorising ``noise tensors'' such as (\ref{eq:Cu4}). 
Consider, for example, the way in which 
the $\sigma $'s were found from the cumulants (\ref{eq:Cu4}) 
(while the latter were actually obtained using the QFT techniques). 
Comparing 
Eqs.\ (\protect\ref{eq:Cu4}) 
to 
Eqs.\ (\protect\ref{eq:FD})%
, we see that the 
(same-time) $\sigma $'s may be specified by postulating 
the characteristic function: 
\begin{eqnarray} % \nonumber 
\label{eq:%
eq:3on%
} 
\Phi (\zeta_1,\zeta_1%
^{\dag}%
,\zeta_2,\zeta_2%
^{\dag} 
) = 
\overline{\hspace{0.1ex}%
{\rm e}%
^{ 
\zeta_1 \sigma_1%
^{\dag} 
+ 
\zeta_1%
^{\dag} 
\sigma_1 + 
\zeta_2 \sigma_2%
^{\dag} 
+ 
\zeta_2%
^{\dag} 
\sigma_2 
}%
\hspace{0.1ex}} 
= 
{\rm e}%
^{ 
- \kappa \zeta _1%
^{\dag 2}%
\zeta _2/8 
- \kappa \zeta _1^2\zeta _2%
^{\dag}%
/8 
} 
. 
\end{eqnarray}%
Although real-valued noises certainly cannot exist which 
satisfy this definition~\cite{Risken}, 
they are easily constructed as complex noises. 
To this end, consider a complex 
Hubbard-Stratonovich transformation~\cite{HST}: 
($x,y$ are arbitrary numbers) 
\begin{eqnarray} % \nonumber 
\label{eq:%
eq:HSTC%
} 
{\rm e}%
^{x y} 
= 
\displaystyle\int \frac{d^2\xi }{\pi} \, 
{\rm e}%
^{%
x\xi + y\xi^*-|\xi|^2 
} 
= 
\overline{\hspace{0.1ex}%
{\rm e}%
^{%
x \xi + y\xi^*%
}%
\hspace{0.1ex}} 
\ \ \ \  
\left[  
xy 
\stackrel{\xi}{\Longrightarrow}%
x \xi + y\xi^* 
\right] . 
\end{eqnarray}%
Here, $\xi $ is 
a standardised complex Gaussian 
noise, with the probability density 
$%
{\rm e}%
^{%
-|\xi |^2%
}%
/{\pi}$. 
The formula in square brackets defines a convenient shorthand; 
using it, we may write: 
\begin{eqnarray} % \nonumber % \eqlabel{} 
- \kappa \zeta _1%
^{\dag 2}%
\zeta _2/8 
\stackrel{%
\xi _2%
}{\Longrightarrow} 
\zeta _1%
^{\dag} 
q \xi_2 + \zeta _1%
^{\dag}%
\zeta _2 p%
^{\dag} 
\xi _2^* 
\stackrel{%
\xi _1%
^{\dag}%
}{\Longrightarrow} 
\zeta _1 
^{\dag} 
q\xi _2 
+ \zeta _1 
^{\dag} 
s\,\xi _1^{\dag *}\sqrt{p^{\dag} \xi _2^{*}} 
+ \zeta _2 r^{\dag} \xi^{\dag} _1 \sqrt{p^{\dag} \xi _2^{*}} 
. 
\end{eqnarray}%
We have thus 
recovered the expressions for the $\sigma _1,\sigma _2%
^{\dag}%
$ 
pair; the $\sigma _1%
^{\dag}%
,\sigma _2$ pair is derived similarly. 

The $\sigma $ noises successfully mimic genuine (non-classical) 
third-order noises, given that averages involving 
their complex conjugates never occur. 
The latter is indeed the case for equations (\ref{eq:FD}). 
With the complex conjugates included, 
we find nonzero cumulants of arbitrary 
order, exactly as expected for non-Gaussian 
statistics. 
Note that the necessity of eliminating complex-conjugate noises 
is exactly the reason why introducing various nonclassical 
noises requires a doubling of the phase space; 
this is no different from 
the positive-P representation. 

In order for equations (\ref{eq:FD}) 
to match the mapping (\ref{eq:HFWM}), the 
values of cumulants mixing 
the increments in (\ref{eq:FD}) 
with their complex-conjugates are irrelevant, 
but from a practical perspective they 
do in fact turn out to be 
very relevant as they affect the sampling noise. 
Consequently the sampling errors can be minimised by 
using the freedom in the definitions of the noises. 
In the present case minimisation of the 
quantity $ 
\overline{\left|\sigma_{1}\right|^{2}}+ 
\overline{\left|\sigma_{1}%
^{\dag}%
\right|^{2}}+\chi\left( 
\overline{\left|\sigma_{2}\right|^{2}}+ 
\overline{\left|\sigma_{2}%
^{\dag}%
\right|^{2}} 
\right) 
$ 
has a noticeably beneficial effect on 
the numerical integration, with 
$\chi$ being a free parameter which 
can be used to 
redistribute noise between the two modes. 
Noting that 
$\overline{|\xi|}=\sqrt{\pi}$, we find the minimum at 
\begin{eqnarray} 
p=p^{\dag} = \frac{\kappa^{1/3}}{4\left(\chi\pi\right)^{1/6}} 
\, , 
\ \ 
s = s%
^{\dag} 
= \chi ^{1/4}, 
\label{eq:%
balance%
} 
\end{eqnarray} 
which via (\ref{eq:Pars}) also fixes the values of $r$'s 
and $q$'s. 
%******************************************* 
\section{Numerical Results} 
%******************************************* 
By numerical experiments we found that, for values of 
$\kappa=\gamma_{j}=1$ and $\epsilon=1.5\epsilon_{c}$, where 
$\epsilon_{c}=\gamma^{2}/\kappa$, a value of $\chi=0.33$ gave the most 
stable results. We should note here that we are working in a very 
strong-interaction regime because this is where it is easiest to see 
the differences between the positive-P and truncated Wigner results. 
This regime is however not necessarily unphysical~\cite{Jevon}. 
We begin our integration with initial 
conditions taken as the 
symmetry-broken 
semiclassical steady state solutions above threshold~\cite{DFW}, 
with $\alpha$ positive. The true physical average 
is however zero, as $\alpha$ may also 
be negative. 
Hence, starting with this initial condition, we see a 
decay of the value of the quadrature average 
$\langle X_{a}\rangle=\overline{\alpha+\alpha^{\dag}}$ 
due to quantum tunneling 
from positive to negative values. As stated above, it is well known that 
the positive-P and truncated Wigner representations give different 
predictions for this tunnelling. 

\begin{figure} 
\begin{center} 
\epsfxsize=0.707\columnwidth 
\epsfbox{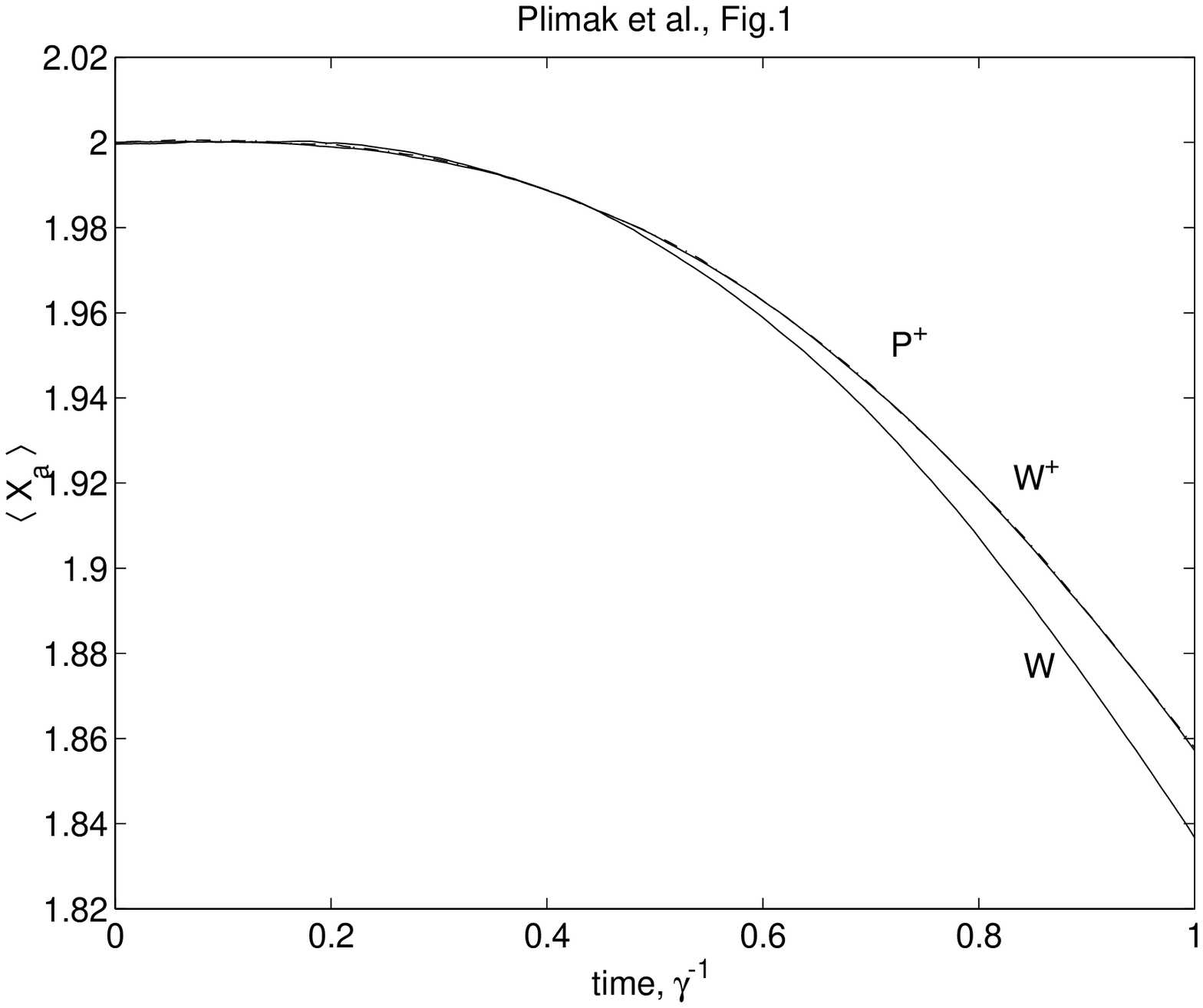} 
\end{center} 
\caption{Predictions of the positive-W (upper solid line), 
positive-P (dashed line) and truncated 
Wigner (solid line) representations for quadrature relaxation in the OPO. 
We can see that the positive-P and positive-W solutions are almost 
indistinguishable.} 
\label{fig:compare} 
\end{figure}

We have numerically integrated the equations of motion in the three 
representations, using a standard Euler technique. We find excellent 
agreement between the positive-P and positive-W results, as can be 
seen in 
Fig.~\ref{fig:compare}, which shows the short time results 
for the quadrature relaxation in the OPO. The positive-W was averaged 
over $1.2\times 10^7$ trajectories, with $2.7\times 10^6$ for the positive-P, 
and 
$1.7\times 10^6$ for the truncated Wigner. This was sufficient to ensure 
excellent convergence over the range plotted. 
Although 
the positive-W 
eventually falls victim to enormous sampling errors, 
where it converges, it 
reproduces the positive-P results almost exactly. The 
truncated Wigner makes noticeably different predictions. 

In summary, we have developed and described a computational method for 
numerical 
modelling of processes which result in generalised Fokker-Planck 
equations with third-order derivatives. Although we cannot define a 
continuous limit of our method as a stochastic process, this is not 
operationally important as interesting systems requiring 
representation with third order noises are likely to be treated 
numerically. We have successfully 
demonstrated our method 
for the example of quantum tunneling 
in the OPO, 
where the neglect of third-order terms in the Wigner 
representation is known to give erroneous results. The success with 
this method gives confidence that 
the technique may be used to model processes where a 
P-representation may require third-order derivatives. The real importance 
of our method is that it 
may be used to extend the use of stochastic integration via 
the phase-space representations beyond the field of quantum optics, 
allowing the deeply quantum aspects of a wider range of systems to be 
investigated. 

%******************************************* 
\stars This research was supported by the Marsden Fund of the Royal Society of New Zealand, 
the New Zealand Foundation for Research, Science and Technology (UFRJ0001),  the 
Israeli Science Foundation and the 
Deutsche Forshungsgemeinshaft. 
%******************************************* 
%\end{document} 

%******************************************* 
%*******************************************
\end{document}